\documentclass[aps,prl,twocolumn,superscriptaddress,longbibliography,nobibnotes]{revtex4-2}
\usepackage{diagbox}
\usepackage{amsmath}
\usepackage{amsfonts}
\usepackage{physics}
\usepackage[space]{grffile}
\usepackage{graphicx} 
\usepackage{amssymb}
\usepackage{upgreek}
\usepackage{textgreek}
\usepackage{lineno}
\usepackage{natbib}
\usepackage{braket}
\usepackage{float}
\usepackage{color}
\usepackage{blindtext}
\usepackage{comment}
\usepackage{bm}
\setcitestyle{super} 

\usepackage{xr-hyper}
\usepackage{hyperref}
\usepackage{xcolor}
\hypersetup{colorlinks,breaklinks,
            urlcolor=[rgb]{0,0,0.64},
            linkcolor=[rgb]{0,0,0.64},
            citecolor=[rgb]{0,0,0.64},
            filecolor=[rgb]{0,0,0.64}}

\usepackage[T1]{fontenc}

\makeatletter
\renewcommand{\thefigure}{\textbf{\@arabic\c@figure}}
\makeatother

\newcommand{\MIT}{Massachusetts Institute of Technology, Department of Physics, Cambridge, Massachusetts 02139, USA.}
\newcommand{\StanfordAP}{Department of Applied Physics, Stanford University, Stanford, California 94305, USA.}
\newcommand{\StanfordPhys}{Departments of Physics, Stanford University, Stanford, California 94305, USA.}

\newcommand{\TDL}{Tsung-Dao Lee Institute, School of Physics and Astronomy, Shanghai Jiao Tong University, Shanghai 200240, China.}

\newcommand{\ICQM}{International Center for Quantum Materials, School of Physics, Peking University, Beijing 100871, China.}
\newcommand{\SLAC}{SLAC National Accelerator Laboratory, Menlo Park, CA 94025, USA.}

\newcommand{\BAQIS}{Beijing Academy of Quantum Information Sciences, Beijing 100913, China.}
\newcommand{\Cornell}{CHESS, Cornell University, Ithaca, New York 14853, USA.}

\newcommand{\SSRL}{Stanford Synchrotron Radiation Lightsource, SLAC National Accelerator Laboratory, Menlo Park, California 94025, USA.}
\newcommand{\APS}{Advanced Photon Source, Argonne National Laboratory, Lemont, Illinois 60439, USA.}
\newcommand{\APSMSD}{Materials Science Division, Argonne National Laboratory, Lemont, Illinois 60439, USA.}
\newcommand{\TUT}{School of Materials Science and Engineering, Tianjin University of Technology, Tianjin 300384, China.}

\begin{document}

\title{
Room-temperature multistage metastability in a moir\'{e} superstructure}


\author{B.~Q.~Lv}
\thanks{These authors contributed equally: B.Q.L., Y.S., and A.Z.}
\affiliation{\TDL}
\affiliation{\MIT}
\author{Yifan~Su}
\thanks{These authors contributed equally: B.Q.L., Y.S., and A.Z.}
\affiliation{\MIT}
\author{Alfred~Zong}
\thanks{These authors contributed equally: B.Q.L., Y.S., and A.Z.}
\affiliation{\MIT}
\affiliation{\StanfordPhys}
\affiliation{\StanfordAP}
\author{Karna Morey}
\affiliation{\MIT}
\author{Bryan~T.~Fichera}
\affiliation{\MIT}
\author{Qiaomei~Liu}
\affiliation{\ICQM}
\author{Dong~Wu}
\affiliation{\BAQIS}
\author{Yongchang~Ma}
\affiliation{\TUT}
\author{Dupeng~Zhang}
\affiliation{\TDL}
\author{Faran Zhou}
\affiliation{\APS}
\author{Makoto~Hashimoto}
\affiliation{\SSRL}
\author{Dong-Hui Lu}
\affiliation{\SSRL}
\author{Donald~A.~Walko}
\affiliation{\APS}
\author{Haidan~Wen}
\affiliation{\APS}
\affiliation{\APSMSD}
\author{Jiarui~Li}
\affiliation{\StanfordAP}
\affiliation{\SLAC}
\author{Suchismita Sarker}
\affiliation{\Cornell}
\author{Jacob P. C. Ruff}
\affiliation{\Cornell}
\author{N.~L.~Wang}
\affiliation{\TDL}
\author{Nuh~Gedik}
\email[Correspondence to: ]{gedik@mit.edu}
\affiliation{\MIT}

\begin{abstract}
  Metastability is fundamental not only to phase ordering and transitions, but also to a broad range of modern technologies, from memory devices to metallic glasses. 
In condensed-matter physics, charge density waves (CDWs) offer versatile platforms for accessing metastable states due to their sensitivity to external stimuli. However, most metastable CDW states are stabilized only at low temperatures, limiting their practical utility. In this study, we report the observation of electrically driven, room-temperature, nonvolatile metastable states in the bulk form of EuTe$_\text{4}$, a recently discovered compound that hosts an innate moir\'{e} superlattice characterized by the stacking of incommensurate monolayer and bilayer CDWs. 
Systematic transport measurements reveal discrete resistivity plateaus and strong electric-field sensitivity, with a large number of metastable states readily induced across a wide temperature window within a giant hysteresis loop, making them well-suited for high-temperature, multi-bit memory applications. By integrating photoemission spectroscopy, diffraction, and \textit{in-situ} transport measurements, we uncover that these metastable states do not stem from conventional mechanisms such as the emergence of new ordered phases or changes in incommensurate periodicity. Instead, they are characterized by a suppression of the original CDW amplitude and a reduction in correlation length, pointing to a unique electric-field-induced switching of out-of-plane CDW phases in the moir\'{e} superstructure. Our findings not only provide critical insights into metastable phenomena in moir\'{e} systems with stacked electronic orders but also establish EuTe$_\text{4}$ as a promising platform for developing room-temperature, multi-bit memory devices.
\end{abstract}

\date{\today}

\maketitle

\section{Introduction}

In many-body systems, the interplay among various degrees of freedom, such as orbital, charge, lattice, and spin, gives rise to a rich array of novel phases, each offering significant potential for advancements in material science and technology\cite{Keimer2017,Cava2021}. The precise control and manipulation of such phases through electric fields is central to the development of modern electronic devices, as exemplified by the recent progress in solid-state resistive memories\cite{Nirmal2024,Loizos2024,Guo2020}. 
Materials exhibiting first-order phase transitions are widely considered among the most promising candidates for such electronic devices, owing to their inherent tendency for competing states\cite{Roch2020,Guillou2018,Dewey2024}. Among these, charge density wave (CDW) materials have garnered particular attention, offering several distinct advantages. First, they represent one of the most prevalent orders in quasi-low-dimensional systems, and different CDW states across a first-order transition are highly accessible to a wide range of experimental probes, including transport, angle-resolved photoemission spectroscopy (ARPES), scanning tunneling microscopy (STM), and diffraction techniques, among others \cite{Lv2022UnconventionalWave,Tomic2009,Sweetland1990}. Second, CDW states are highly responsive to electric fields, which can directly influence the motion of charged ions and electrons within the crystal lattice, enabling field-driven phase transitions\cite{Gruner2018DensitySolids,BROCK1993}. Moreover, the transition to metastable CDW states is typically accompanied by apparent changes in resistance due to the changing of energy gaps near the $E_\text{F}$, making it easier for reading\cite{Monceau1976,Fleming1979}. A prominent example is 1$T$-TaS$_2$, where a hidden, nonvolatile CDW state with fast switching speeds and substantial resistance changes have been demonstrated using optical or electrical pulses\cite{Zong2018UltrafastWave,Vaskivskyi2015,Ravnik2018,sto2014,masaro2015,Ma2016}. Despite these advancements, the search for ideal materials remains challenging. Specifically, materials with a large thermal hysteresis width, an essential parameter that defines the operating temperature range of CDW-based devices, are still scarce. 

\begin{figure*}[htb!]
\centering
\includegraphics[width=0.9\textwidth]{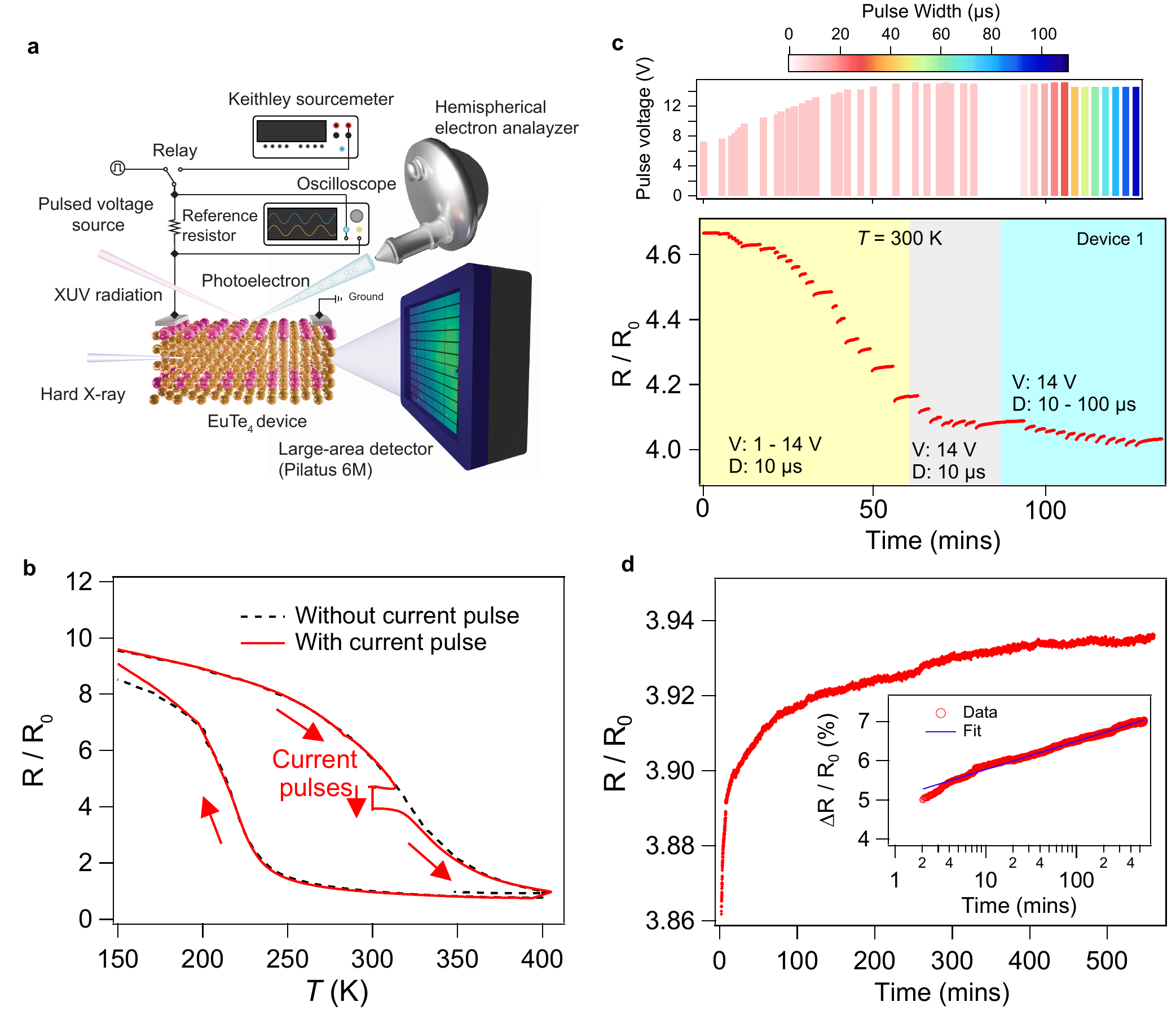}
\caption{\textbf{EuTe$_\text{4}$ device characterized by multi-messenger probes.} \textbf{a},~Schematics of experimental setups of X-ray diffraction (XRD) and angle-resolved photoemission spectroscopy (ARPES) integrated with \textit{in-situ} pulsed voltage excitation and time-resolved transport measurement circuits. \textbf{b},~Dashed black line: Temperature-dependent electrical resistance of a EuTe$_4$ device normalized by the value at 400~K ($R_0 = 208~\Omega$), featuring a giant thermal hysteresis spanning the entire temperature window. Red solid line: Similar temperature scan with a series of pulsed voltage excitations (specified in \textbf{c}) at 300~K in the heating branch, featuring an instant decrease in resistance, which can be recovered via thermal annealing. \textbf{c},~Normalized resistance of a EuTe$_4$ device with a series of applied pulsed voltage excitations as indicated in the top panel. The series of excitations contains three parts of controlled experiments. In the first regime (yellow), we applied 10~$\upmu$s pulses with increasing voltage, featuring a drop in resistance. The amount of drop positively scales with the applied voltage. In the second regime (purple), we applied pulses with a constant voltage (14~V) and duration (10~$\upmu$s), showing saturation behavior. In the third regime (blue), we applied pulses with fixed voltage (14~V) and increasing pulse duration from 10~$\upmu$s to 100~$\upmu$s, demonstrating a further decrease in resistance with the application of longer pulses. \textbf{d},~Evolving resistance after pulse excitation, featuring a slow recovery on a timescale of approximately 9~hrs following the pulse excitations. Inset:~same data plotted on a log timescale, featuring a logarithmic recovery as a function of time.}
\label{fig:Fig1}
\end{figure*}

Recently, EuTe$_4$, which hosts a rare quasi-two-dimensional (quasi-2D) CDW moir\'{e} superstructure, has been discovered and has attracted increasing attention due to several peculiarities. Structurally, EuTe$_4$ belongs to the $R$Te$_n$ family~—~a class of extensively studied layered CDW materials \cite{Yumigeta2021AdvancesSynthesis,Ru2008,Shin2010,Shin2005,Shin2008}. 
As the only known tetratelluride in this family, EuTe$_4$ exhibits a unique crystal structure consisting of alternating monolayer and bilayer Te square-net sheets. This distinctive structure leads to the coexistence of monolayer and bilayer CDWs, characterized by two incommensurate modulation vectors, $\mathbf{q}_1 = 0.644(5)\mathbf{b}^*$ and $\mathbf{q}_2 = 0.678(5)\mathbf{b}^* +0.5\mathbf{c}^*$, respectively. The intricate interplay between these two stacking CDWs gives rise to a large moir\'{e} superlattice, with an in-plane periodicity reaching up to 13.6~nm. This opens an exciting frontier for exploring  innate moir\'{e} physics that emerges from intrinsically stacked incommensurate orders. In particular, due to distinct out-of-plane arrangements of the incommensurate CDWs \cite{Lv2025LargeWaves}, the resulting  moir\'{e} superstructure exhibits a pronounced thermal hysteresis over a broad temperature range of 100--500~K --- a record among crystalline solids \cite{Wu2019LayeredSheets,Lv2022UnconventionalWave}. Such an exceptionally wide hysteresis window offers a long-sought platform for investigating metastable states triggered by optical or electrical stimuli at or above room temperature. 
Very recently, pioneering transport and optical measurements demonstrated optical and electrical switching of hidden, nonvolatile CDW states in EuTe$_4$ flakes, marking the beginning of understanding and controlling metastable behaviors in such a room-temperature moiré superstructure\cite{Liu2024,venturini2024}.

\begin{figure*}[htb]
\centering
\includegraphics[width=\textwidth]{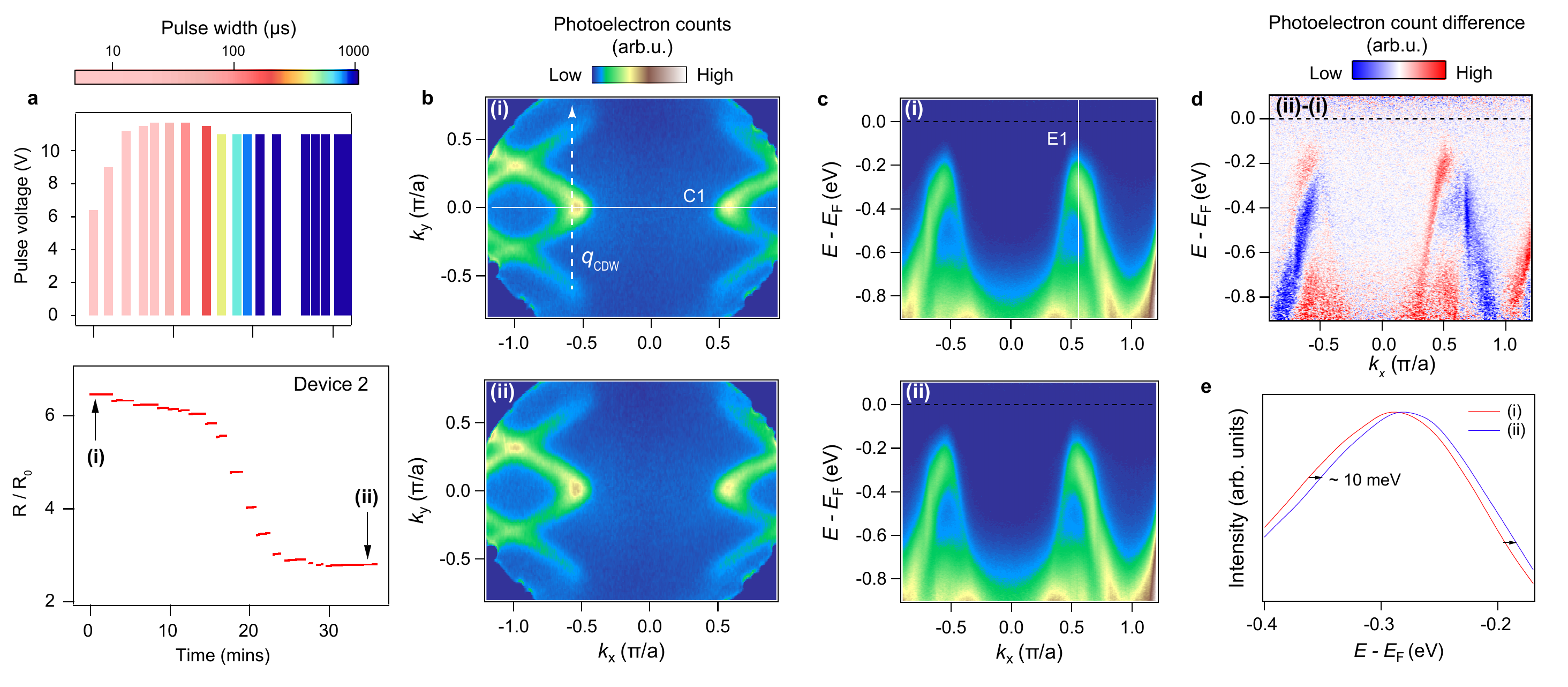}
\caption{\textbf{Electronic structure of EuTe$_\text{4}$ at 300 K upon pulsed voltage excitation.} \textbf{a},~Normalized resistance as a function of time measured \textit{in-situ} in the ARPES measurement chamber. Each discontinuity in resistance corresponds to a pulsed voltage excitation of 11~V. The sequence of voltage pulses applied is indicated in the top panel. The resistance saturates after applying 16 pulses. \textbf{b},\textbf{c},~Constant energy contour at $-0.2$~eV (\textbf{b}) and energy-momentum dispersion cut along the $\Gamma$-X direction (\textbf{c}) of EuTe$_4$ device before (i) and after (ii) series of pulsed voltage excitations. The timestamps (i) and (ii) correspond to those marked with arrows in panel \textbf{a} The dashed arrow denotes the CDW wavevectors averaged under the finite momentum resolution. \textbf{d},\textbf{e},~Difference map (\textbf{d}) of energy-momentum dispersion cuts between the spectrum before and after pulsed voltage excitation $(ii)-(i)$, together with energy distribution curves (EDC) (\textbf{e}),  highlighting the upward shift of the band leading edge after electrical excitation.}
\label{fig:Fig2}
\end{figure*}

In this study, we combine \textit{in-situ} time-resolved transport measurements with ARPES and X-ray diffraction (XRD) measurements to systematically investigate electrically excited metastable states in pristine bulk crystals of EuTe$_4$. We provide direct evidence of voltage-induced nonvolatile metastable CDW states at and above room temperature. Uniquely, we find that multiple metastable states can be accessed depending on the number of voltage pulses applied. These states are associated with variations in the amplitude, shape, and correlation length of the CDWs, while the in-plane incommensurate CDW wavevectors remain preserved. We interpret the metastable states as a result of the formation of multiple types of CDW domains, differentiated by the relative CDW phases among distinct quasi-2D Te layers. This phenomenon is unique to this particular stacking CDW  moir\'{e} system with intermediate interlayer coupling and does not occur in purely quasi-2D or strongly coupled 3D systems\cite{Yumigeta2021AdvancesSynthesis,Wilson2024}. Our findings provide new insights into the nature of metastable states in EuTe$_4$, establishing a robust and versatile platform for the design of high-temperature CDW memory devices.

\section{Results}
We start by characterizing the EuTe$_4$ device using systematic transport measurement with a circuit design demonstrated in Fig.~\ref{fig:Fig1}a (see details in Methods and Supplementary Information). Upon the application of a voltage pulse voltage, we observed successive electrical switching operations. Figures \ref{fig:Fig1}b and \ref{fig:Fig1}c illustrate a representative example at 300~K within the heating branch of the thermal hysteresis loop, where a series of in-plane electrical pulses with progressively increasing voltages drives the system into multi-step metastable states, each characterized by abrupt resistance jumps. This electrically driven switching is particularly intriguing as it satisfies many criteria for an ideal memory device\cite{Kim2012}. First, the metastable states are nonvolatile and exhibit an exceptionally slow, quasi-logarithmic relaxation process. As shown in Fig.~\ref{fig:Fig1}d, the resistance change remains under 7$\%$ over a measured time window of approximately nine hours. Second, while these metastable states are nonvolatile, they are fully reversible through a thermal erase procedure, as demonstrated in Fig.~\ref{fig:Fig1}b. Third, these states can be triggered over a wide temperature range within the heating branch of the hysteresis loop, particularly between 300~K and 400~K (Fig.~\ref{fig:Fig1}c and S1), which is highly desirable for applications in high-temperature storage devices\cite{Pradhan2024}. Fourth, the switching process is fast, at least within the intrinsic pulse duration, i.e., on the order of microseconds. Finally, the metastable states and their accompanying sharp resistance jumps are very sensitive to the amplitude of the applied electric field, facilitating easy writing and reading in terms of memory operation.  
These distinctive features highlight EuTe$_4$ as a practical and promising platform for the development of long-sought, wide-temperature-range, non-volatile, and multi-bit resistive memories. We note that these metastable states exhibit a non-negligible dependence on pulse duration (see Fig.~\ref{fig:Fig1}c). In particular, a pronounced dependence is observed for longer pulses (see Fig.~\ref{fig:Fig4}b). This observation suggests that, in addition to electric-field effects, thermal contributions associated with Joule heating cannot be ruled out in the formation of these metastable states (see the Supplementary Information for detailed analysis of Joule heating).

Having established the existence of electrically driven metastable states, we now turn to the underlying mechanisms and macroscopic characteristics of these states beyond their resistance signatures. As previously noted, EuTe$_4$ hosts two distinct, interacting CDWs within a single unit cell. These CDWs are differentiated by their modulation wave vectors, giving rise to a large innate moir\'{e} superstructure. Prior ARPES and XRD measurements have shown that the observed hysteresis does not stem from conventional incommensurate-to-commensurate lock-in transitions. Instead, it originates from unconventional switching between different three-dimensional configurations of the moir\'{e} superstructure\cite{Lv2022UnconventionalWave,Lv2024CoexistenceSemiconductor,Lv2025LargeWaves}. Recent high-resolution XRD studies have further revealed that the coexisting monolayer and bilayer CDWs are jointly locked to the lattice via the relationship $q_1+2q_2=2b$, where $q_1$ and $q_2$ are the in-plane components of wavevectors for monolayer and bilayer CDWs, respectively, which accounts for the extraordinary robustness of $q_1$ and $q_2$ against temperature changes\cite{Lv2025LargeWaves} as well as optical excitation on an energy scale of around 1~eV\cite{Oh2025JCCDW}. These findings underscore that the interplay between the coexisting monolayer and bilayer CDWs is central to understanding the voltage-induced metastable states. To this end, we employed an \textit{in-situ} setup that integrates transport measurements with ARPES or XRD characterization, as illustrated in Fig.~\ref{fig:Fig1}a. This integrated approach allows direct access to the CDW order parameters and thus a comprehensive exploration of the electronic and structural responses, offering deeper insights into the mechanisms governing the formation and stability of the electrically driven metastable states in EuTe$_4$.

Figure~\ref{fig:Fig2}a displays the in-situ resistance measured in the ARPES chamber at the 300~K heating branch. Consistent with the results presented in Fig.~\ref{fig:Fig1}c, the application of voltage pulses drives the system into multiple lower-resistance states, which persist until the arrival of the next pulse, emphasizing their non-volatile nature. To probe the evolution of the electronic structure, ARPES measurements were conducted both before and after the application of voltage pulses. Figures \ref{fig:Fig2}~b--e summarize the measured constant-energy contours (0.2~eV below Fermi Surface), electronic band dispersion along the $\Gamma$--X direction, and energy distribution curves (EDCs) before and after pulse injections, as indicated by points (i) and (ii) in Fig.~\ref{fig:Fig2}a. A comparison of the electronic structures between these two states reveals that the overall electronic structure remains unchanged before and after the voltage pulse. This observation confirms that the voltage pulses do not generate new CDW modulations and that both the metastable and original states arise from the same underlying CDW distortions. On the other hand, the significant resistance reductions observed in Fig.~\ref{fig:Fig2}a suggest suppression of the CDW after the pulse injection. The CDW suppression can be quantified through ARPES measurements as the band gap opened near Fermi surface, which can be directly measured by ARPES, is proportional to the amplitude of the CDW order parameter, and thus the amplitude of periodic lattice distortion induced by CDW\cite{Gruner1988,Lv2019Angle-resolvedMaterials}. Figure~\ref{fig:Fig2}e summarizes the EDCs at the E1 point, revealing a clear upward shift of approximately 10~meV after the voltage injection. The upward shift was also clearly visible in the difference map of energy-momentum dispersion presented in Fig.~\ref{fig:Fig2}d [$(ii)-(i)$]. This shift indicates a weakening of CDW strength in the metastable states. To quantify the evolution of the valence band top under electric pulses with increasing electric field and pulse duration, we systematically tracked the leading-edge positions ($E_L$) of the EDCs. The main results are summarized in the Supplementary Information. A gradual upward shift of is observed, which correlates well with the concurrent change in resistance. This consistency points towards a progressive suppression of the CDW strength in the metastable phase.

\begin{figure*}[htb!]
\centering
\includegraphics[width=0.8\textwidth]{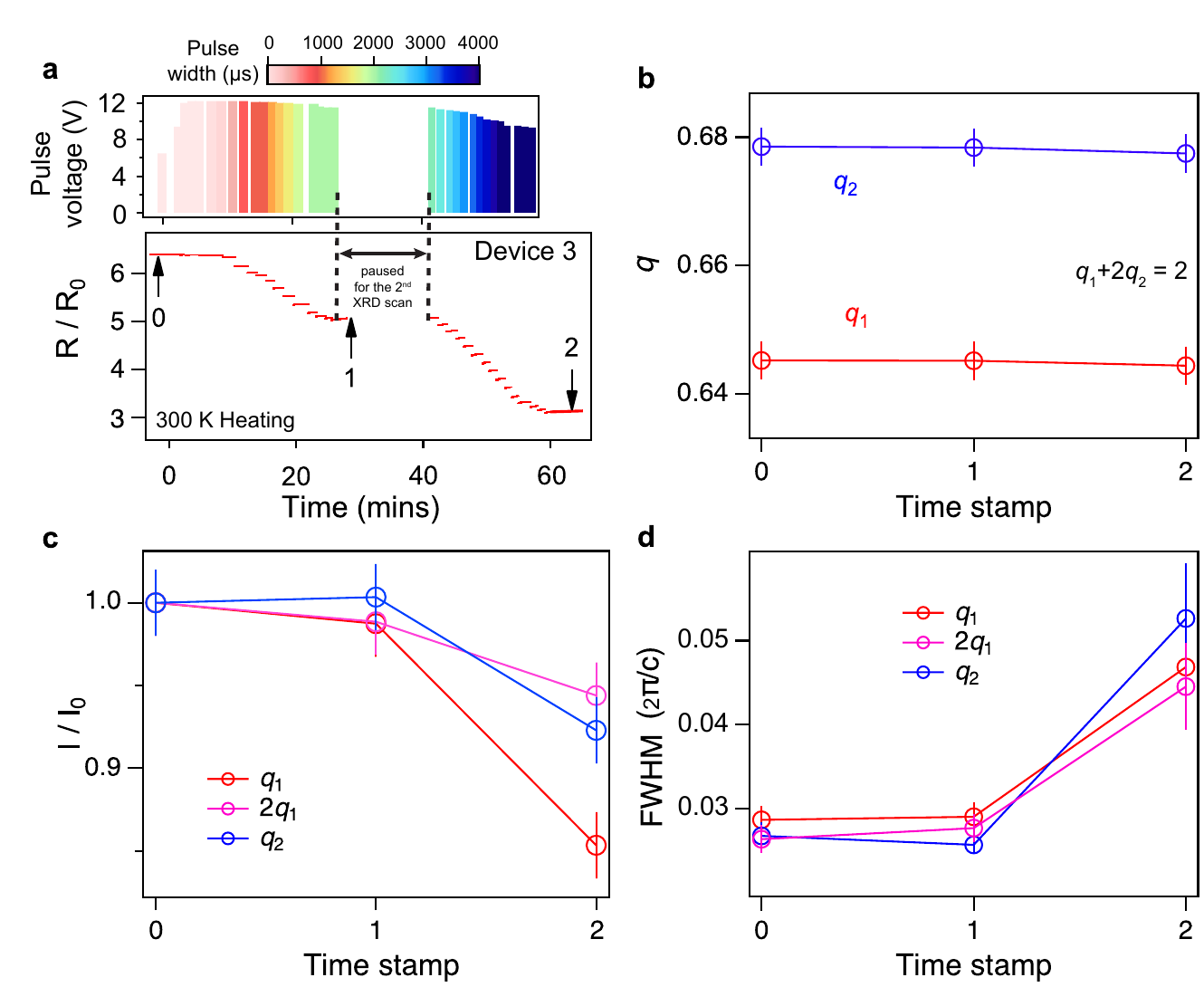}
\caption{\textbf{Voltage-induced suppression of CDW distortion probed by X-ray diffraction.} \textbf{a},~Normalized resistance as a function of time measured \textit{in-situ} at the CHEXS XRD endstation. The sequence of applied voltage pulses is indicated in the top panel. \textbf{b},~In-plane component of monolayer ($q_1$) and bilayer ($q_2$) CDW wavevectors before and after pulsed voltage excitation, showing robust jointly commensurate relation without noticeable change upon excitations. The error bar represents instrumental momentum resolution. \textbf{c},\textbf{d},~Normalized intensity (\textbf{c}) and full width at half maximum (FWHM) (\textbf{d}) of monolayer ($q_1$) and bilayer ($q_2$) CDW satellite peaks as well as second-order satellite of monolayer CDW ($2q_1$). All data are averaged over CDW peaks in over 10 Brillouin zones in the reciprocal space mapping to boost the signal-to-noise-ratio, given the consistency across different Brillouin Zones. The error bars represent the statistical uncertainty derived
from the fitting procedure. }
\label{fig:Fig3}
\end{figure*}

To further verify the features of CDWs upon electrical excitation, we turn to structural characterization and deploy XRD reciprocal space mapping. XRD measurements not only characterize the amplitude of order parameters but also precisely determine the CDW modulation wavevector and the phase coherence. XRD results measured at 300~K (heating branch) together with in-situ transport measurements are shown in Fig.~\ref{fig:Fig3}. The main characteristics of metastable states can be summarized with five key observations. First, the in-plane modulation q-vectors for both the monolayer and bilayer CDWs remain unchanged after pulse injection, as shown in Fig.~\ref{fig:Fig3}b. This robustness indicates the preservation of the joint locking relationship between the monolayer and bilayer CDWs under both temperature and electrical pulse perturbations\cite{Lv2025LargeWaves}. Second, in general, both the monolayer and bilayer CDWs exhibit a reduced strength after voltage injection, as evidenced by the clear suppression of the integrated intensity of the superlattice peaks shown in Fig.~\ref{fig:Fig3}c. Upon closer inspection, however, we find that the superlattice peak intensity changes only weakly under relatively small electric pulses, even though the resistivity already exhibits a sizable change. This asynchronous behavior suggests that the observed reduction in resistivity arises from a combination of domain formation and weakening of the CDW amplitude. On the other hand, we note that the CDW gap measured by ARPES exhibits an approximately linear response to voltage injection (see the Supplementary Information). We attribute this behavior to the probing depth of the measurement techniques. ARPES is a surface-sensitive probe, while XRD is bulk-sensitive; consequently, the voltage-induced responses of the surface and bulk CDW orders may differ. Third, the monolayer and bilayer CDWs exhibit distinct responses to electrical fields. As shown in Fig.~\ref{fig:Fig3}c, the monolayer CDW superlattice peak exhibits a larger decrease in integrated intensity after pulse perturbation than the bilayer CDW superlattice peaks. This observation, consistent across all order of diffraction peaks, suggests that the monolayer CDW is more susceptible to electrical perturbations compared to the bilayer CDW. It should be noted that such differential renormalization has also been observed under optical pulse perturbation\cite{Lv2024CoexistenceSemiconductor}. In both cases, the distinct renormalization behavior can be attributed to the different gap sizes of monolayer and bilayer CDWs. Fourth, after pulse injection, the first- and second-order diffraction peaks in all Brillouin zones exhibit distinct behaviors (Fig.~\ref{fig:Fig3}c). As previous XRD studies revealed, the monolayer CDW potentially manifests as a non-sinusoidal lattice distortion, producing clear diffraction intensities at the second and third harmonics of the superlattice peaks\cite{Lv2025LargeWaves,Lv2022UnconventionalWave}. The relatively larger decrease in the intensity of the first-order superlattice peaks in Fig.~\ref{fig:Fig3}c thus indicates a potential change towards non-sinusoidal shape in monolayer CDW under electrical fields. Fifth, the broadening of the full width at half maximum (FWHM) of CDW superlattice peaks along the $c$-axis after electrical excitation (Fig.~\ref{fig:Fig3}d) indicates a reduction in the out-of-plane correlation length of the metastable CDW states. This suggests that electrical excitation induces disorder and potentially multiple domains, contributing to the observed sharp resistance drop. The discrete resistance platueaus also indicates a finite number of CDW domains being switched. These observations reveal that while the in-plane periodicity and joint lock-in nature of the CDWs are preserved, the metastable states are characterized by notable changes in the strength, shape, and out-of-plane correlation length of the monolayer and bilayer CDWs. These findings provide critical insights into the nature of metastable states and their tunability under electrical fields.

\begin{figure*}[htb!]
\centering
\includegraphics[width=0.8\textwidth]{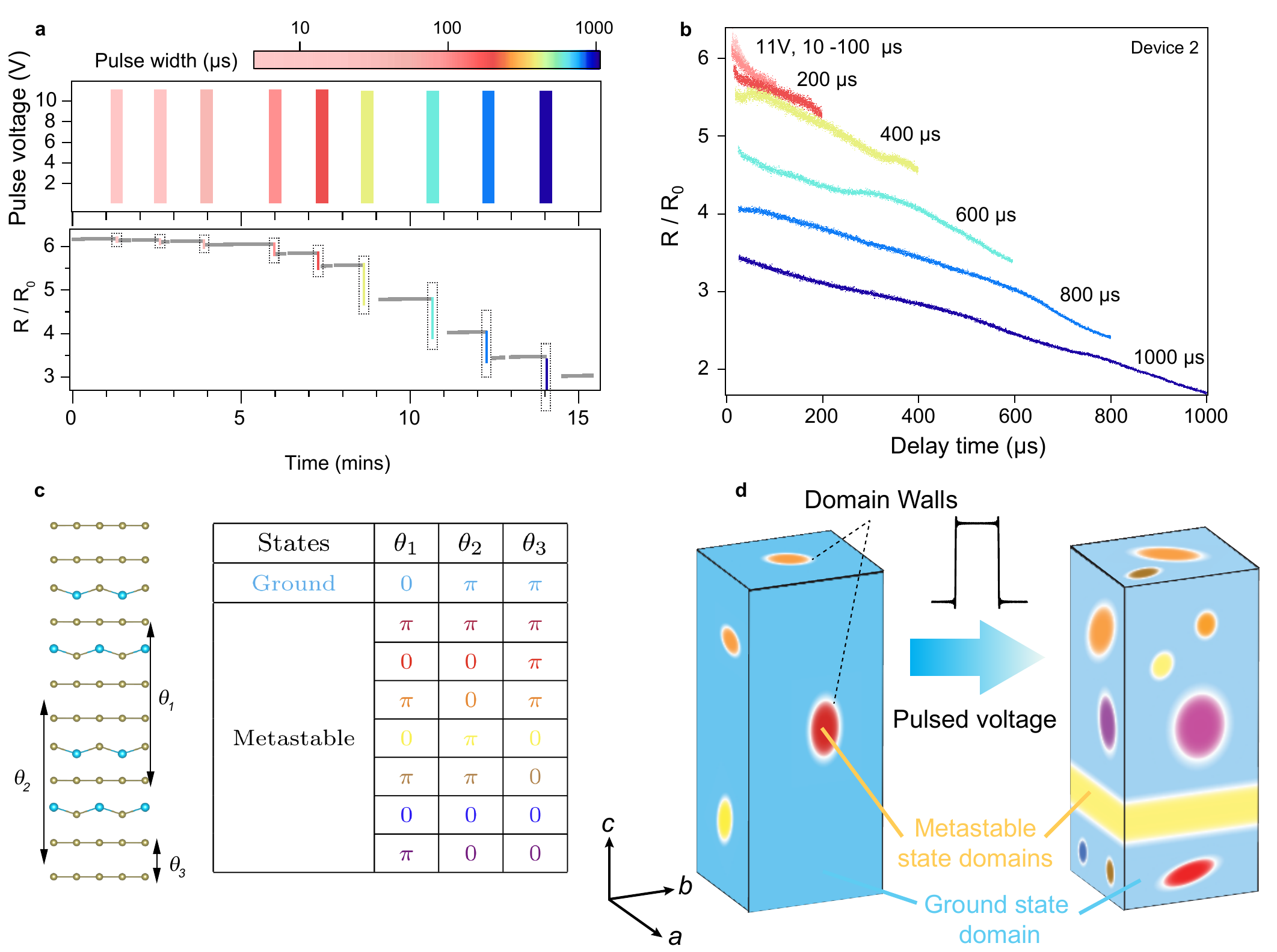}
\caption{\textbf{Time-resolved resistance measurement and voltage-induced metastable CDW domain formation.} \textbf{a},\textbf{b},~Time evolution of resistance during the application of 11~V electrical pulses of varying durations, ranging from 10~$\upmu$s to 1~ms, as indicated in the top panel of \textbf{a}. For all cases with different applied pulse durations, the decrease in resistance persists for the entire span of pulse duration, suggesting a continuous field-driven metastable state in CDW orders. The measurement is performed on the same device as in the ARPES measurement.  \textbf{c},~List of all eight possible CDW states defined by relative phases between Te monolayer sheets in adjacent unit cells ($\theta_1$), between Te bilayer sheets in adjacent unit cells ($\theta_2$), and within Te sheets in each unit cell ($\theta_3$). \textbf{d},~Schematics of voltage-induced metastable CDW states. The cuboids indicate the domain dynamics in real space, labeled with crystalline axes. The area with cyan color represents the ground state, and areas with other colors represent metastable CDW domains, where different colors indicate different metastable phases with distinct relative phases between the Te layers, as color-coded in the table in \textbf{c}. The thin white boundaries indicate domain walls between distinct CDW domains. Upon pulsed electrical excitation, the ground state CDW amplitude is suppressed, and metastable CDW domains emerge.}
\label{fig:Fig4}
\end{figure*}

Thus far, we have focused on the macroscopic manifestations of the metastable states after the application of electrical pulses. To achieve a comprehensive understanding of metastable behaviors, it is essential to probe the dynamic responses of the CDWs during the pulse injection. Key issues to investigate include how the original CDW evolves into the metastable state and whether the original CDW is completely quenched during the pulse injection process. However, it is very challenging to measure the transient resistance of the material during the pulse injection. To address these questions, we developed a time-resolved resistance measurement system that utilizes a high-bandwidth oscilloscope. Figure~\ref{fig:Fig4}a illustrates the circuit design, where a reference resistor is employed to derive the transient resistance using the voltage divider equation
\begin{equation}
    R_s = \frac{V_2}{V_1-V_2}R_\text{ref},
\end{equation}
where $R_\text{ref} = 47~\Omega$ is the reference resistor and $V_1$ and $V_2$ are the voltages on the two ends of the reference resistor measured by two channels in the oscilloscope. Figure~\ref{fig:Fig4}b summarizes the time evolution of resistance under electrical pulses of varying durations, ranging from 10~$\upmu$s to 1~ms. The most important observation is a gradual decrease in the resistance over time, indicating that the original CDW state is not fully melted, even under electrical pulses as high as 11~V and lasting up to 1~ms. This gradual decrease persists until the end of the pulse, highlighting the critical role of the applied electric field in driving the system toward metastable CDW states. These findings provide valuable insights into the dynamic evolution of CDWs under electrical perturbations. They suggest that the transition to metastable states is more likely a continuous, field-driven process rather than an abrupt or complete destruction of the original CDW configuration. However, we recognize that other possibilities, such as thermal quenching or topological defect formation, cannot be fully excluded and need further studies.\cite{Zong2019b,Orenstein2023}

\section{Discussion}
Summarizing the characteristics of the metastable states observed during and after pulse injection, we propose that the metastable CDW states arise from the continuous formation of multiple CDW domains with different relative phases. Previous XRD measurements revealed that, in the ground state, the out-of-plane periods of the monolayer and bilayer CDWs are $c$ and 2$c$, respectively\cite{Lv2025LargeWaves}. These correspond to relative CDW phases of $\theta_1=0$ for adjacent Te monolayers and $\theta_2=\pi$ for adjacent bilayers, defining the ground-state configuration as ($\theta_1$,$\theta_2$) = (0, $\pi$) (see Fig.~\ref{fig:Fig4}c). Beyond the ground state, the interplay and coupling between monolayer and bilayer CDWs give rise to three additional classes of metastable states, distinguished by the relative phases ($\theta_1$,$\theta_2$): (0,0), ($\pi$,$\pi$), and ($\pi$,0). The energy differences between these metastable configurations and the ground state are approximately 10~meV according to density functional theory calculation\cite{Dong2026}, which is comparable to thermal fluctuations at 300~K\cite{Lv2025LargeWaves}, enabling the emergence of multiple metastable CDW moir\'{e} domains. The observed anomalously large hysteresis can be attributed to the hysteretic temperature-dependent evolution of domain sizes associated with these metastable states\cite{Lv2025LargeWaves}. Given the quasi-2D nature of the Te layers, the in-plane couplings are significantly stronger than the out-of-plane couplings. As a result, the application of electric fields predominantly modifies the out-of-plane periodicity of the monolayer and bilayer CDWs, while the in-plane periodicity remains preserved, as evidenced by our in-situ ARPES and XRD measurements in Fig.~\ref{fig:Fig2} and Fig.~\ref{fig:Fig3}. Under sufficiently strong electric fields, it is also plausible that the relative phase ($\theta_3$) within the Te bilayer could be altered. By restricting $\theta_1$, $\theta_2$, and $\theta_3$ to values of 0 or $\pi$, a total of eight distinct configurations can arise, as listed in Fig.~\ref{fig:Fig4}c. As previously studied in time-resolved diffraction studies, The observed weaker CDW strength and reduced correlation lengths suggest that the gradual evolution of CDW moir\'{e} domain sizes\cite{Zong2019b}, with varying 3D configurations, as illustrated in Fig.~\ref{fig:Fig4}d, underpins the multi-step nature of the metastable states. However, in spite of the large number of nearly degenerate metastable states that make the specific CDW state distribution probabilistic upon excitation, the resistivity change remains reproducible upon excitation as the average CDW amplitude and correlation length predominantly depend on the pulse energy of the excitation pulse.

It is also important to clarify that the present work is the first to investigate metastable phenomena in bulk EuTe$_4$. This is in sharp contrast to \citet{Liu2024} and \citet{venturini2024}, which focus on micro-flakes with thicknesses that are approximately three orders of magnitude smaller than those of bulk crystals. As a result, the metastable behaviors observed in bulk crystals and thin flakes differ markedly. First, in the cooling branch, bulk crystals exhibit little change in resistivity under voltage driving (see Supplementary Information, Sec.~III). In contrast, in thin flakes, a single voltage pulse of 2~V or an ultrafast optical pulse with fluence exceeding can drive the system into hidden high-resistance metastable states \cite{Liu2024,venturini2024}.
Second, in the heating branch, bulk crystals subjected to a sequence of voltage pulses with increasing amplitude (up to 12~V) evolve into multiple metastable states characterized by substantially reduced resistivity (exceeding a 50\% drop, see Fig.~\ref{fig:Fig2}a). In contrast, in thin flakes, voltage pulses below 2~V induce metastable states with lower resistivity, whereas voltages exceeding 2~V drive the system into hidden high-resistance states.
These differences are not incidental but instead directly reflect the underlying physics. As discussed previously, owing to the quasi-two-dimensional nature of EuTe$_4$, electric fields predominantly modulate the out-of-plane CDW stacking. In thin flakes, the limited number of CDW units along the $c$ axis allows the entire system to be driven into a single metastable configuration, resulting in hidden high-resistance states. In bulk crystals, however, the vastly larger number of CDW units makes such global alignment unlikely, naturally leading to multiple metastable states and the absence of hidden high-resistance phases.

By simultaneously examining changes in transport properties and lattice and electronic structures under a series of electrical pulses, we identified several distinctive characteristics of the metastability in EuTe$_4$: (1)~the metastable CDW states exhibit nonvolatile behavior with exceptionally long lifetimes, (2)~they are highly sensitive to electric field strength rather than pulse duration, (3)~multiple metastable states are observed, and they can be readily distinguished by their distinct resistance values, and (4)~they can be triggered over a broad temperature range within the hysteresis loop. These findings position EuTe$_4$ as an ideal platform for the development of wide-temperature range multi-bit data storage devices. By tracking the band structure and CDW satellite peaks before and after electrical pulse excitation, we confirmed that the Fermi surface topology and in-plane periodicity remain unaltered, safeguarded by the joint lock-in nature of the stacking monolayer and bilayer CDWs. However, the observed weakening of the CDW strength, along with the broadening of the FWHM of satellite peaks, suggests that the metastable states arise from transitions between distinct 3D configurations of the CDW moir\'{e} superstructure. These findings deepen our understanding of the phase evolution of CDWs under electrical perturbations and provide a compelling framework for interpreting metastable phenomena in low-dimensional systems with stacked electronic orders. Furthermore, they lay the groundwork for the fabrication of next-generation nonvolatile memory devices capable of wide-temperature-range operation, leveraging transitions between out-of-plane phases of moir\'{e} superstructure with large intrinsic differences in electrical resistance.

\section{Methods}
\subsection{Sample preparation}
EuTe$_4$ single crystals were synthesized via a solid-state reaction with Te as the flux. Stoichiometric Eu lumps (99.999\%) and Te granules (99.999\%) were mixed with a ratio of approximately 1:15. The total weighted starting materials were sealed in an evacuated fused silica tube under high vacuum ($10^{-5}$ mbar) followed by heating at $850^\circ$C for two days in a muffle furnace. The furnace was slowly cooled to $415^\circ$C over 100 hr, held at this temperature for one week, and then decanted using a centrifuge \cite{Wu2019LayeredSheets}. Unlike $R$Te$_3$, EuTe$_4$ single crystals are stable under ambient conditions. They are planar-shaped with dark and mirror-like surfaces, where the surface area is up to 1~mm$^2$ and the thickness up to 0.2~mm. Synthesized bulk crystals are cut into rectangular strips and fixed onto various substrates with silver epoxy in preparation for device fabrication. Two-contact transport device is fabricated by attaching silver wires on the two ends of the strip-like samples with silver epoxy. The device can be further cleaved on the top surface for XRD and ARPES measurements. 

\subsection{Pulsed voltage source and transport measurement}
To provide voltage pulses, we use a standard Agilent 33220A single-channel function generator to generate pulses of adjustable width between 500 ns and 100 ms at a repetition frequency of 0.3 Hz. An operational amplifier (Analog Devices AD811) in series with current buffers (BUF634F) is used to amplify the signals from the function generator for sufficient voltage and current. The output of the current buffer is then connected to a 47~$\Omega$ reference resistor in series with the sample. For $\upmu$s-scale time-resolved measurements, the voltage relative to the ground is probed before and after the reference resistor by two channels in an oscilloscope (RIGOL DS2202A), therefore, the resistance of the sample can be computed by using the voltage divider equation. To measure the resistance to high accuracy seconds to hours after applying pulsed voltage while putting minimal current through the sample in the steady state (especially to maintain cryogenic temperatures), we construct an isolated circuit with an ohmmeter (Keithley 2400) in series with the sample. An electromechanical relay is used to switch between the two circuits mentioned above. More detailed discussions on the circuit design can be found in the Supplementary Information.

\subsection{X-ray reciprocal space mapping}

X-ray reciprocal space mapping experiments were carried out at the QM2 beamline at the Cornell High Energy Synchrotron Source (CHESS) \cite{chess}. The incident X-ray energy of 29~keV was selected using a double-bounce diamond monochromator. A stream of cold-flowing nitrogen gas was used to cool the sample. The diffraction experiment was conducted in reflection geometry using a 6-megapixel photon-counting pixel-array detector with a silicon sensor layer (Pilatus 6M). Data were collected in 90$^\circ$ sample rotations with a step size of 0.1$^\circ$. After data acquisition, all images are stacked together to form 3D data labeled with three diffraction angles. A least-squares fitting is performed on all identified peaks to generate an orientation matrix that maps the data from the diffraction angle space to the $(H,K,L)$ space.

\subsection{Angle-resolved photoemission spectroscopy}
High-resolution ARPES measurements were performed at Beamline 5-2 of the Stanford Synchrotron Radiation Lightsource (SSRL) with a Scienta DA30 electron analyzer. The photon energy used was 25~eV, and the combined (beamline and analyzer) experimental energy resolution was 6~meV and 16~meV for energy-momentum cuts and Fermi surface maps, respectively. The angular resolution of the DA30 analyzer was 0.2$^\circ$. The beam spot had an approximate cross-sectional size of 50~$\upmu$m by 100~$\upmu$m. The chemical potential was determined from the spectra of a polycrystalline gold that was electrically connected to the measured sample. Fresh surfaces were obtained by cleaving EuTe$_4$ crystals \textit{in-situ} in ultrahigh vacuum, where the base pressure was maintained below $2.2\times10^{-11}$ torr for all temperatures investigated. All constant-energy maps shown in the main text and Supplementary Material have an intensity integration window of 20meV centered at the specified energy.

\section{Acknowledgments}
The authors thank Riccardo~Comin, Anshul~Kogar, Honglie~Ning, Kyoung~Hun~Oh, and Darius~Shi for helpful discussions. The authors thank Evan Zalys-Geller for helpful scientific discussions regarding designs of the pulsed voltage circuitry. The work at MIT was supported by the U.S. Department of Energy, the BES DMSE (data collection and analysis), and the Gordon and Betty Moore Foundation's EPiQS Initiative grant GBMF9459 (manuscript writing). Research conducted at the Center for High-Energy X-ray Science (CHEXS) is supported by the National Science Foundation (BIO, ENG, and MPS Directorates) under award DMR-2342336. Research conducted at Stanford Synchrotron Radiation Lightsource (SSRL) at SLAC National Accelerator Laboratory is supported by the U.S. DOE, Office of Science, Office of Basic Energy Sciences under contract no.~DE-AC02-76SF00515. B.L. acknowledges support from the National Natural Science Foundation of China (Grant Nos.~12374063, 92565305), the Ministry of Science and Technology of China (Grant No.~2023YFA1407400), and the Shanghai Natural Science Fund for Original Exploration Program (Grant No.~23ZR1479900). A.Z. acknowledges the support from the U.S. Department of Energy, Office of Basic Energy Sciences under award No.~DE-SC0026202 (data analysis and manuscript writing). K.M. acknowledges support of the MIT Undergraduate Research Opportunities Program (UROP). N.L.W. acknowledges support from the National Natural Science Foundation of China (Grant No.~12488201), and the National Key Research and Development Program of China (Grant Nos.~2024YFA1408700, 2022YFA1403901). D.W. acknowledges support from the National Key Research and Development Program of China (Grant No.~2024YFA1408700). F.Z. and H.W. acknowledge the support for part of the x-ray diffraction measurements by the U.S. Department of Energy, Office of Science, Basic Energy Sciences, Materials Sciences and Engineering Division, under award No.~DE-SC0012509. Part of this research used resources of the Advanced Photon Source, a U.S. Department of Energy (DOE) Office of Science user facility operated for the DOE Office of Science by Argonne National Laboratory under Contract No.~DE-AC02-06CH11357. 

\section{Author contributions}
B.Q.L., Y.S., and A.Z. conceived the study. B.Q.L., Y.S., A.Z., K.M., B.F., Y.M., D.Z., and J.L. performed transport measurements. B.Q.L., Y.S., A.Z., S.S., and J.P.C.R. performed the X-ray diffraction measurements. B.Q.L., A.Z., D.L., and M.H. performed the ARPES measurements. Q.L., D.W., and N.L.W. synthesized, characterized, and prepared the EuTe$_4$ crystals. K.M., B.Q.L., and B.F. designed the transport set-up. S.S. and J.P.C.R. maintained and set up the synchrotron end-station at CHESS. F.Z., D.A.W., and H.W. assisted with the X-ray diffraction and \textit{in situ} transport measurements at the Advanced Photon Source. B.Q.L., Y.S., and A.Z. analyzed the data with the help of Y.M. and D.W. B.Q.L., Y.S., and A.Z. wrote the manuscript with critical input from N.G. and all other authors. The work was supervised by N.G.

\section{Competing Interests}
The authors declare no competing interests.

\end{document}